# Thermal Dephasing in the Laughlin Quasiparticle Interferometer: Chiral Luttinger Liquid Behavior

F. E. Camino, W. Zhou and V. J. Goldman
*Department of Physics, Stony Brook University, Stony Brook, New York 11794-3800, USA*

We report experimental temperature dependence of the amplitude of Aharonov-Bohm oscillations in the Laughlin quasiparticle interferometer. The results fit very well the thermal dephasing dependence predicted for a $g = 1/3$ chiral Luttinger liquid interferometer, and are clearly distinct from the activated behavior observed in single-particle resonant tunneling and Coulomb blockade devices. The small deviation from the zero-bias theory seen below 20 mK indicates yet unrecognized source of experimental decoherence, not included in theory.

The familiar Fermi liquid picture of interacting electrons as Landau quasielectrons in one-to-one correspondence to the undressed electrons breaks down in 1D. The charge and spin excitations in a variety of physical systems demand entirely different effective theories describing 1D electrons, the Tomonaga-Luttinger liquids. [1-4] A 2D electron system in a quantizing magnetic field forms 1D edge conducting channels, the chiral Luttinger liquids ($\chi$LL), [5] the direction of circulation being determined by the field. The elementary charged excitations, Laughlin quasiparticles (LQPs), [6] of a fractional quantum Hall (FQH) fluid have fractional electric charge and anyonic braiding statistics. [6-9] While the interior of the FQH fluid is gapped, and the LQPs are localized by residual disorder, there is no gap for charged edge excitations, and their low-energy dynamics is described by the $\chi$LL theories. Theoretical proposals to probe the $\chi$LL behavior of the FQH edges by electron and quasiparticle tunneling [10-15] stimulated much experimental work. Dramatic power-law current-voltage characteristics have been observed for an external electron tunneling into a FQH edge. [16] Most internal quasiparticle tunneling experiments, in single constrictions and in quantum antidots, [17,18] however run into the difficulty that the differences between the predicted $\chi$LL and the Fermi liquid behavior are surprisingly small in these experimental settings (for not entirely understood reasons), even at $T \sim 10^{-3} E_F$, the Fermi energy. We are not aware of any experimental study of a LQP interferometer prior to Ref. 9 and this work.

We report here experiments on thermal dephasing of the Aharonov-Bohm oscillations in the novel LQP interferometer, where quasiparticles of the 1/3 FQH fluid execute a closed path around an island of the 2/5 fluid. Qualitatively, the experimental results follow a thermal dephasing dependence expected for an electron interferometer, and show clear distinction from the activated behavior observed in resonant tunneling and Coulomb blockade devices, both in the $\chi$LL and the Fermi liquid regimes. The data fit very well the $\chi$LL dependence predicted for a two point-contact LQP interferometer. [13] The fit yields a value of the chiral edge excitation velocity, obtained for the first time for a continuous FQH edge excitation spectrum. At the lowest experimental $T \leq 20$ mK, the small deviation from the zero-bias theoretical $\chi$LL dependence can be fit extremely well by including a finite bias, improving the overall fit. However, the value of the bias obtained in the fit is three times higher than the total of the known experimental sources, perhaps pointing to yet unrecognized source of experimental decoherence not included in theory.

The LQP interferometer sample, illustrated in Fig. 1(a), was fabricated as described formerly. [9] The chiral edge channels form at the periphery of the undepleted 2D electrons, following the constant electron density equipotentials around the etch trenches. Here we focus on the situation when a QH filling $f = 1/3$ annulus surrounds an island of the $f_I = 2/5$ FQH fluid, as shown schematically in Fig. 1(a). We are confident that the current is transported by the $f = 1/3$ fluid because the Hall resistance is quantized to $R_{XY} = 3h/e^2$, see Fig. 1(b). In this regime, we observe the Aharonov-Bohm conductance oscillations as a function of *B*, with the flux period $\Delta\Phi = 5h/e$, Fig. 2(a), as reported before. [9] The island center electron density is only 4% less than the well-known 2D bulk density, and thus has essentially the same QH filling. The island $f_I = 2/5$ is further confirmed by the ratio of the flux to the backgate oscillation periods, which is proportional to $1/f_I$, and was calibrated in the integer QH regime.

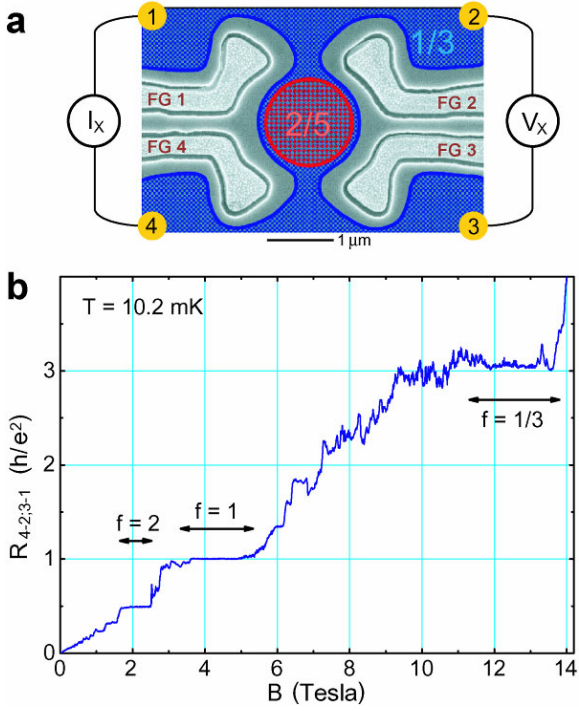

**Fig. 1.** (a) SEM micrograph of a typical GaAs/AlGaAs heterostructure device. The front gates (FG) in shallow etch trenches define the central island of 2D electrons ~300 nm below the surface. The main depletion potential is produced by the etch trenches, the FG's are used for fine-tuning the two wide constrictions for symmetry. The four numbered Ohmic contacts are in fact at the corners of a 4×4 mm sample. The filling 2/5 island is surrounded by the 1/3 FQH fluid, which forms the Aharonov-Bohm ring. (b) Four-terminal Hall resistance $R_{XY} \equiv V_{4-2}/I_{3-1}$ is determined by the QH filling $f$ in the constrictions, giving definitive values. The fine structure is due to quantum interference effects, including the Aharonov-Bohm oscillations as a function of magnetic flux through the electron island.

The temperature dependence of the oscillations for $10.2 \leq T \leq 141$ mK is shown in Fig. 2(b), where the directly measured four-terminal $R_{XX} = V_X/I_X$. The ~12.9 kΩ background results from the quantized $R_{XX} = R_{XY}(\frac{1}{3}) - R_{XY}(\frac{2}{5}) = \frac{1}{2} h/e^2$. These particular $R_{XX}(T)$ data were obtained continually over ~70 hour period (following a ~100 hour sample stabilization period), which demonstrates the stability of the Aharonov-Bohm conductance oscillations in this FQH regime.

In order to be able to quantitatively compare the experiment to a theory, we analyze these raw data as follows. We determine the amplitude of each of the several regular oscillations, $\delta R_{XX}(T)$, at each temperature. The oscillatory conductance $\delta G$ is calculated [5,17] from the directly measured $\delta R_{XX}(T)$ and the quantized value of the Hall resistance $R_{XY} = 3h/e^2$ as $\delta G = \delta R_{XX}/(R_{XY}^2 - \delta R_{XX} R_{XY})$ for $\delta R_{XX} \ll R_{XY}$. We then normalize $\delta G(T)$ to the average of the two lowest temperatures (10.2 and 11.9 mK) for each particular oscillation, and take the average of $\delta G(T)$ for six thus normalized oscillations, to reduce the experimental uncertainty. The normalized conductance amplitude $G_A(T)$ data are shown in Fig. 3 on the linear and semi-log scales. The experimental $G_A(T)$ varies by 31× when $T$ varies by 14×.

It is apparent in the semi-log plot that the higher-$T$ oscillation amplitude follows $G_A(T) \propto \exp(-T/T_0)$, the dependence expected for thermal dephasing of a quantum interference signal, with a surprisingly small $T_0 \approx 28$ mK, yet the oscillations persist up to ~140 mK. This is qualitatively distinct from an activated behavior, where $G_A(T) \propto \exp(-T_0/T)$, observable only for $T < T_0$. The experimental $G_A(T)$-dependence is also clearly different from that observed in the two well-known Fermi liquid tunneling phenomena: resonant tunneling in quantum dots [19] and antidots, [20] and in the Coulomb blockade devices, [21] where the conductance oscillations fade away well before $T = T_0$.

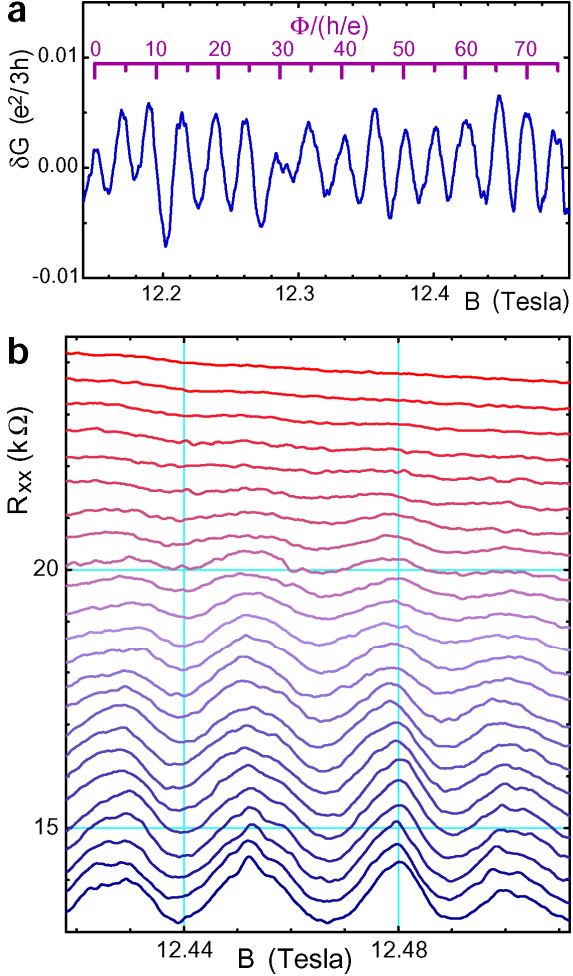

**Fig. 2.** Aharonov-Bohm oscillations of Laughlin quasiparticles. (a) Magnetic flux through the island period $\Delta\Phi = 5h/e$ corresponds to creation of ten $e/5$ quasiparticles in the island ($h/e$ induces two LQPs in the 2/5 fluid). The $e/3$ LQP consecutive orbits around the island are quantized so that the total Berry phase acquired in circling the island (sum of the $(e/3\hbar)\Delta\Phi$ Aharonov-Bohm and the anyonic contributions) is an integer multiple of $2\pi$. (b) Evolution of the Aharonov-Bohm oscillations in the $10.2 \leq T \leq 141$ mK temperature range. The successive traces are offset vertically by 0.4 k$\Omega$. Oscillatory conductance $\delta G$ is calculated from the directly measured $R_{XX}$ as described in the text. The data in (a) and (b) were obtained on different cooldowns of the sample.

We compare our experimental $G_A(T)$ data to several theoretical predictions. Chamon *et al.* have calculated the Aharonov-Bohm oscillatory conductance expected for a two point-contact interferometer formed by $\chi$LL edge channels. [13] Their geometry is similar to ours, the most apparent difference is that they consider only one QH filling in the interferometer device. Their calculated amplitude of oscillations is proportional to a function of two reduced variables $H_g(y_1, y_2)$, where $y_1 = 2\pi\omega_J/\omega_0$ and $y_2 = \omega_J/2\pi T$. The "Josephson frequency" for the charge $q$ LQPs is $\omega_J = qV_H/\hbar$, and the "oscillatory frequency" is $\omega_0 = 4\pi u/C$, where $u$ is the edge excitation propagation velocity, and $C$ is the interferometer circumference. The explicit expression for the function $H_g(y_1, y_2)$ for the primary Laughlin QH states, characterized by the Luttinger exponent $g$ ($g = 1/3$ for the $f = 1/3$ FQH fluid), is given in Eqs. 21, 22 in Ref. 13.

We first fit the low bias $\omega_J \to 0$ theoretical dependence to the experimental data. Since we normalize the experimental $G_A(T)$ to $G_A(11\,\text{mK})$, the theoretical curves [that is, $H_g(y_1, y_2)$] are not normalized to 1 as $T \to 0$. The best fit, shown by the solid line in Fig. 3, gives the $e/3$ LQP effective tunneling amplitude $|\Gamma_{eff}|^2 = \Gamma_1\Gamma_2^* + \Gamma_1^*\Gamma_2 = 0.008$ (from the absolute value of $G_A(11\,\text{mK})$, before normalization), and the LQP edge velocity $u = 1.4\times10^4$ m/s (from $\hbar\omega_0 = 334$ mK). The fit is generally very good; a small but systematic discrepancy is seen in that the experimental points are below the theoretical curve at low $T$, and above in the intermediate range $35 < T < 80$ mK. The small experimental $|\Gamma_{eff}|^2 \ll 1$ validates our use of the theoretical expression for $H_g(y_1, y_2)$ obtained to the lowest order in the tunneling amplitudes. Additionally, we obtain the effective radial confining electric field at the chemical potential $\mathcal{E} = uB = 1.7 \times 10^5$ V/m. We use the $f = 1/3$ edge channel circumference $C = 4.0$ μm, corresponding to the radius of 570 nm obtained for this approximately circular device. [9] The values of the LQP edge velocity and the confining electric field are comparable (~2× smaller) to the values reported for a quantum antidot. [17,20] Apparently, the screening of the bare confining potential is more efficient in this LQP interferometer device than in the smaller radius quantum antidots.

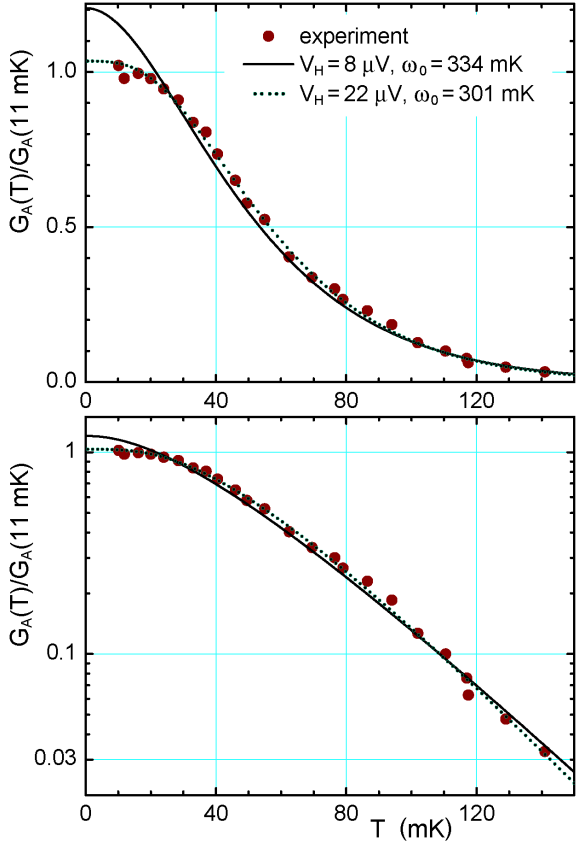

**Fig. 3.** Temperature dependence of the amplitude of oscillations in the LQP interferometer. The fits use the g = 1/3 chiral Luttinger Liquid theory of Chamon et al., Ref. 13, with "oscillation frequency" $\omega_0$ as the fitting parameter. The $V_H = 8$ μV (total of the applied bias and the "noise") finite bias fit is virtually indistinguishable from the $V_H = 0$ fit on this scale. Increasing bias to $V_H = 22$ μV makes the χLL fit exceptionally good, but such a high value of $V_H$ is not feasible; this is understood as indicating an additional experimental source of decoherence, not included in theory. The high-$T$ thermal dephasing asymptotic behavior $G_A(T) \propto \exp(-T/T_0)$ is evident in the semi-log plot.

Including the effects of a finite Hall voltage, that is non-zero $\omega_J$, makes the fit extremely good, shown by the dotted line in Fig. 3. The best-fit value of $\omega_J$, however, yields the Hall voltage of 22 μV, some 2.8 times higher than the combined contributions of the known applied measurement bias and the integrated electromagnetic "noise" incident on the sample. An estimate of the noise voltage [17] is ~2 μV rms, the applied Hall voltage was 7.2 μV rms, for a combined ~7.5 μV rms. Experimentally, reducing the applied bias by a factor of 2 results in a 2-3% increase of the oscillation amplitude at 10.2 mK. Thus the combined effect of the applied bias and the electromagnetic noise is not sufficient to account for the low-$T$ discrepancy. A possible explanation of the discrepancy (the too high best-fit value of $\omega_J$) is an additional source of experimental decoherence, not included in theory. For example, tunneling LQP's induce dissipative Faraday currents in the metal front gates, which could contribute such additional decoherence. [22] Even if we attribute all the discrepancy to decoherence, still the combined "noise"-induced and the unknown-source decoherence of the interference signal is only 8% at the lowest experimental $T$.

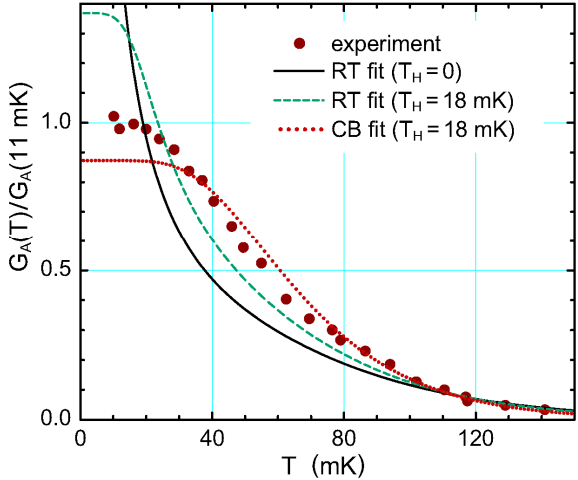

**Fig. 4.** Two-parameter best fits to oscillatory behavior expected for single-particle resonant tunneling (RT) and Coulomb blockade (CB) devices. $T_H$ is the "electron heating" temperature. The RT fit is very sensitive to the value of $T_H$ because of the $1/T$ Fermi liquid prefactor. The CB oscillation amplitude saturates at ~32 mK, higher than $T_H$. The two CB curves ($T_H = 0$ and 18 mK) are virtually indistinguishable on this scale. $T_H = 18$ mK was reported in Ref. 17 for a smaller quantum antidot device, where electron heating is expected to be stronger than in the larger LQP interferometer.

The LQP interferometer $G_A(T)$ dependence is very different from that observed in resonant tunneling in quantum dots [19] and antidots, [17,20] where the single-particle energy spacing $\Delta E \gg T$, and in the Coulomb blockade devices, [21] where $\Delta E, \Gamma \ll T$. In resonant tunneling, a single tunneling peak is well described by $G(X,T) = G_0/T \cosh^2(X/T)$, where $X$ is the resonance tuning parameter, even in the FQH regime. The single peak conductance $G_P(T) \propto 1/T$. For the classical Coulomb blockade regime, not expected in this nearly open geometry, [12] a single tunneling peak is given by $G(X,T) = G_0 X/T \sinh(X/T)$. The single peak conductance $G_P(T) \approx const$ because the single-charge tunneling can proceed via many levels within an energy interval $\propto T$, canceling the $1/T$ Fermi liquid prefactor. In both regimes, the individual $G(X,T)$ tunneling peaks overlap at higher temperatures, resulting in a non-universal amplitude of oscillations dependence $G_A(T)$, since it depends on the level spacing $\Delta E$ or the charging energy $U_C$. At arbitrary temperatures, the two $G_A(T)$ dependencies can be evaluated numerically, with activation $T_0 \propto \Delta E$ or $\propto U_C$ as a fitting parameter.

Fig. 4 gives the best fit of the experimental interferometer $G_A(T)$ to the resonant tunneling (RT) theory. We show two fits with two free parameters, the peak conductance $G_0 \propto \Gamma$ and the level spacing $\Delta E$: one assuming no electron heating effects, and another including the heating effects, taking the experimental $T_H = 18$ mK, [17] obtained in the limit of bath $T \to 0$. The fit is very sensitive to the value of $T_H$ because of the diverging $1/T$ prefactor. We stress that no physically realistic value of $T_H$ produces a good fit to the LQP interferometer data. Fig. 4 also gives the best two-parameter fit of the experimental interferometer $G_A(T)$ to the Coulomb blockade (CB) theory. The large deviation of the fit at low $T \ll U_C$ is apparent, with the experimental $G_A(T)$ still rising as $T \to 0$, while the theoretical dependence saturates at ~32 mK. This behavior is opposite to that in the RT theory, and is not curable by including electron heating effects. Additionally, the CB $T$-dependence results from the conductance minima rising, while the maxima remain nearly constant, with increasing $T$, a behavior clearly different from that seen in this sample (*cf.* Fig. 2(b)). Both RT and CB fits give the energy scale ~500 mK, surprisingly large for a ~2 μm lithographic diameter device. Thus we conclude that neither the single-particle resonant tunneling nor the classical Coulomb blockade occurs in the LQP interferometer device.

As mentioned above, the experimental apparent high-$T$ thermal dephasing $G_A(T) \propto \exp(-T/T_0)$ has a small $T_0 \approx 28$ mK. As a comparison, in a quantum antidot in the FQH regime, [17] the corresponding activation $T_0 \approx 120$ mK, while the oscillations were observable up to ~80 mK, a lower $T$ than the present ~140 mK. In the quantum dots and antidots in the integer QH regime, similarly, oscillations are observable up to $\sim 0.5 T_0$. To elucidate the restraint the LQP interferometer data impose on a possible alternative theory, we also fit the data to a phenomenological expression $G_A(T) = G_0 (T/T_0)^\alpha [\sinh(T/T_0)]^{-1}$, where power α is an additional parameter. Reasonable fits are obtainable only with $\alpha \approx 1 \pm 0.3$ (and including $T_H$ = 18 mK), so that the strictest restriction on the functional dependence of $G_A(T)$ in the experimentally-accessible regime is placed by the prefactor to the asymptotic behavior. Neither $\alpha = -1$, as for the Fermi liquid resonant tunneling, nor $\alpha = 5$ predicted for the $g = 1/3$ Aharonov-Bohm effect in a strong-coupling limit in a quantum antidot [14] agrees with the experimental LQP interferometer $G_A(T)$ dependence.

The theory of Ref. 13 includes no island with QH filling other than that of the interferometer edge ring itself. Thus it is not presently known whether inclusion of such island should modify their Eq. 22. Such modification seems unlikely, however, on general grounds, since the $e/5$ island LQPs do not participate in current transport, and the interaction of the interfering $e/3$ LQPs with the $e/5$ island LQPs is nonlocal statistical interaction [23] of a topological nature only. Such statistical interaction should not affect the dynamics of the interfering $e/3$ LQPs so long as the number of the island LQPs remains well-defined, that is, at temperatures $T << 3$ K, the $f_I = 2/5$ FQH gap at 12 Tesla.

We thank D. V. Averin for discussions. This work was supported in part by the NSF and by NSA and ARDA through US ARO.